\documentclass{emulateapj}

\usepackage {epsf}
\usepackage {lscape, graphicx}
\usepackage {psfig}
\usepackage {amssymb,amsmath}
\usepackage [figuresleft]{rotating}

\gdef\1054{MS\,1054--03}
\gdef\2053{MS\,2053--04}

\submitted{}

\begin {document}

\title {The detailed Fundamental Plane of two high redshift clusters: \2053\ at z=0.58 and \1054\ at z=0.83}

\author{Stijn Wuyts\altaffilmark{1}, Pieter G. van Dokkum\altaffilmark{2}, Daniel D. Kelson\altaffilmark{3}, Marijn Franx\altaffilmark{1}, and Garth D. Illingworth\altaffilmark{4}}
\altaffiltext{1}{Leiden Observatory, P.O. Box 9513, NL-2300 RA, Leiden, The Netherlands.}
\altaffiltext{2}{Department of Astronomy, Yale University, P.O. Box 208101, New Haven, CT06520-8101.}
\altaffiltext{3}{Carnegie Observatories, 813 Santa Barbara Street, Pasadena, CA 91101.}
\altaffiltext{4}{University of California Observatories/Lick Observatory, University of California, Santa Cruz, Santa Cruz, CA95064.}

\begin {abstract}
We study the fundamental plane relation in high redshift clusters
using a sample of 26 galaxies in \2053\ ($z=0.583$) and 22 galaxies in
\1054\ ($z=0.83$).  The zeropoint and scatter are compared to results
for lower redshift clusters in order to trace evolutionary
effects. Furthermore, our large sample enables us to investigate
correlations between residuals from the fundamental plane and other
characteristics of the galaxies, such as color, $H\beta$ linestrength,
spatial distribution, and mass.  The observed scatter of the
early-type galaxies with $\sigma > 100\ km\ s^{-1}$ around the
fundamental plane is $0.134$ and $0.106$ in $\log r_e$ for \2053\ and
\1054\ respectively.  The residuals from the fundamental plane of
\2053\ are correlated with residuals from the $H\beta - \sigma$
relation, suggesting that stellar populations are playing a role in
shaping the fundamental plane.  The measured evolution in $\log M/L$
is influenced by selection effects, as galaxies with lower M/L in the
Johnson B-band enter a magnitude-limited sample more easily.  When we
select high mass early-type galaxies to avoid this bias, we find $\log
M/L_B \sim -0.47 z$ and a formation redshift $z_{form} \sim 2.95$,
similar to earlier results.
\end {abstract}


\section{INTRODUCTION}

In the local universe early-type galaxies follow a tight scaling
relation, known as the Fundamental Plane (FP).  The relation between
effective radius, central velocity dispersion and surface brightness
$r_e \sim \sigma^{\alpha} I_e^{\beta}$, which is a plane in $(\log
r_e, \log \sigma, \log I_e)$ space, was discovered by
Djorgovski \& Davis (1987) and Dressler et al.\ (1987).  In
combination with the virial theorem
\begin {equation}
M/L \sim \sigma^2 r^{-1} I_e^{-1}
\label {virial.eq}
\end {equation}
the small scatter around the FP implies that, under the assumption of homology, the $M/L$ ratios
of early-type galaxies are well behaved and scale as
\begin {equation}
M/L \sim \sigma^{\alpha/\beta+2} r_e^{-(1+\beta)/\beta}.
\label {MLpred.eq}
\end {equation}
As the $M/L$ ratio increases with ageing of a stellar
population, the FP is a useful tool in research on galaxy formation
and evolution.  Based on a sample of 226 E and S0 galaxies in 10
clusters of galaxies, J$\o$rgensen, Franx,\& Kj$\ae$rgaard (1996,
hereafter JFK96) concluded the local plane in the Johnson B band has
the form:
\begin {equation}
\log r_e = 1.20\log \sigma - 0.83\log I_e + \gamma
\end {equation}
which implies that
\begin {equation}
M/L \sim M^{0.28}r_e^{-0.07}.
\end {equation}
Later studies on intermediate and high redshift clusters of galaxies
used the zeropoint shift of the plane to estimate average formation
redshifts of the stars in early-type galaxies (e.g., Bender et
al. 1998; van Dokkum et al.\ 1998, hereafter vD98; J$\o$rgensen et
al. 1999; Kelson et al.\ 2000c, hereafter K2000; Pahre et al. 2001;
van Dokkum \& Stanford 2003).  The scatter around the plane provides
constraints on the spread in galaxy ages.

The slope of the FP (and other scaling relations) constrains
systematic age trends with mass and other parameters.  Any evolution
of the slope of the FP with redshift implies that the ages of the
stellar populations are correlated with galaxy mass.  For the cluster
CL1358+62 at z = 0.33 K2000 finds that the slope of the FP has not
evolved significantly over the past $\sim 4\ Gyr$.  The sample of 5
bright \2053\ galaxies used by Kelson et al.\ (1997, hereafter K97) to
study the FP at z = 0.583 seemed to agree with the values of
coefficients $\alpha$ and $\beta$ as given by JFK96.  However, the
sample was too small to perform a proper fit.  The same conclusions
were drawn for \1054\ at z = 0.83 based on 6 early-type galaxies
(vD98).  In this paper we investigate the FP of \2053\ and \1054\
early-type galaxies using larger samples spread over a larger range of
distances from the brightest cluster galaxies (BCG).  In Sect.\
\ref{spectroscopy.sec} we discuss the spectroscopy, sample selection
and velocity dispersions.  Imaging and measurement of the structural
parameters is described in Sect.\ \ref{imaging.sec}.  Zeropoint of the
FP with JFK96 coefficients and scatter around the plane are discussed
in Sect.\ \ref{jorg.sec}.  In Sect.\ \ref{correlation.sec} we study
correlations between the residuals from the FP and various other
properties of the galaxies.  Finally the conclusions are summarized in
Sect.\ \ref{conclusion.sec}. VEGA magnitudes are used throughout this
paper.  We use $H_0=70\ km\ s^{-1}\ Mpc^{-1}$, $\Omega_{\Lambda}=0.7$,
and $\Omega_M=0.3$, but note that our results are independent of the
value of the Hubble constant.

\section {SPECTROSCOPY}
\label {spectroscopy.sec}

\subsection {Sample selection and observations}

All spectra used in the FP analysis were obtained with the LRIS
spectrograph (Oke et al.\ 1995) on the 10 m W.M. Keck Telescope.  The
data were obtained in a series of observing runs from 1996 to 2002.
The majority of galaxies in our final sample were selected on the
basis of their spectroscopic redshift and their $I$- or F814W-band
magnitude.  The samples were limited at $I\approx 22$ for all runs;
galaxies with $I\lesssim 21$ were given highest priority in the mask
designs.  The redshift information came from a large $I$-selected
spectroscopic survey of both clusters described in detail in Tran et
al.\ (1999), van Dokkum et al.\ (2000), and Tran (2002).  For the
initial \1054\ observing runs only limited redshift information was
available, and we applied color criteria to select likely cluster
members.  Galaxies having $\Delta(R-I)-0.25 \Delta (B-R)<-0.4$, with
$\Delta(R-I)$ and $\Delta(B-R)$ colors relative to the central galaxy,
were excluded.  The color ranges were chosen such that blue cluster
members were unlikely to be excluded.  The final FP sample of \1054\
contains 19 galaxies that were selected with these mild color
constraints.  No morphological information was used in the selection
process.

For most observations we used the 600 lines\,mm$^{-1}$ grating blazed
at 7500\,\AA; some of our earlier data were taken with the 831
lines\,mm$^{-1}$ grating blazed at 8200\,\AA\ (see vD98).  The
wavelength coverage was typically $\sim 3500$ to $\sim 5400 \AA$ in
the restframe.  Exposure times ranged from 7500\,s to 33400\,s and
from 10500\,s to 22800\,s for the \2053\ and \1054\ galaxies
respectively.  The instrumental resolution was typically
$\sigma_{inst} \sim 40-80\ km\ s^{-1}$ and signal-to-noise ratios
ranged from 20 to 100$\AA^{-1}$ in the observed frame (in the continuum).

A total of 43 galaxies (26 early-type) were observed in \2053.  The
\1054\ sample contained 30 galaxies (14 early-type).  The
morphological classification is described in Sect.\
\ref{classification.sec}.  Early-type galaxies include E, E/S0 and S0
morphologies.

\subsection{Basic reduction}
\label{reduction.sec}

The spectra were reduced using our own software and standard IRAF
software routines (see, e.g. Kelson et al. 2000b).  The wavelength
calibration was performed using the night sky emission lines.  The
typical rms scatter about the fitted dispersion solutions is about
1/15 of a pixel. Since the dispersion is about $1.28\, \AA\
pixel^{-1}$ for the data taken with the $600\ mm^{-1}$ grating (and
$0.92\, \AA\ pixel^{-1}$ for the $831\ mm^{-1}$ data), the rms scatter
is equivalent to velocity errors smaller than $5\ km\ s^{-1}$.

The flat-fielding accuracy is generally better than a percent, on small
scales. The data have not been accurately flux-calibrated so on large
scales, the notion of flat-fielding accuracy is not meaningful.  There tend
to be moderate-scale ($k=100\, \AA^{-1}$) residuals in the flat-fielding that are
the result of a mismatch between the fringing in the flat-fields and the
fringing in the data.  Such inaccuracies in de-fringing the data have no
effect on the velocity dispersions because the spectra are effectively
filtered on those scales (and larger) in the process of matching the
continua of the template and galaxy spectra.

The subtraction of the sky was performed using standard, published, and
well-tested methods: for each galaxy the two-dimensional spectra were first
rectified, and then 1st- or 2nd-order polynomials were fit to the pixels on
both sides of the galaxy, typically excluding the few arcsec where the
galaxy is bright. An iterative, clipping routine was used to reject any
remaining bad pixels from the fit, similar to what IRAF allows the user to
do.  These methods have been discussed elsewhere at great length and we
choose not to bore the reader with very familiar territory.  While more
complicated and precise means of sky subtraction are now available (see
Kelson 2003), these data were analyzed before such methods became
available, and the accuracy of the sky subtraction performed in the
"traditional" way is satisfactory for our purposes.

The slit widths varied slightly with each run, ranging from $0\farcs
90$ to $1\farcs 05$.  The extraction apertures for \2053 and \1054
were 7 CCD rows (or $1\farcs 5$).

\subsection {Velocity dispersions}
\label {dispersion.sec}

Velocity dispersions were measured with a direct fitting method
(Kelson et al.\ 2000b). Details of the fitting procedure are given in
Kelson et al.\ (2003).  Direct fitting methods provide several
advantages over Fourier-based techniques. Most importantly, pixels are
no longer weighted uniformly in the computation of the fitting metric
(in this case $\chi^2$).  As in Kelson et al.\ 2000b, we weight each
pixel by the inverse of the expected noise (photon and electronic read
noise). Furthermore, the spectra that exhibit strong Balmer absorption
had those features given zero weight in the fitting, in order to
minimize the contribution of those features to template mismatch
error, and also to ensure that our estimates of $\sigma$ reflected the
old, underlying stellar populations.  We typically fit the spectra
over a $\sim 1000\, \AA$ wavelength range (in the restframe).  We used
a range of template stars, from G5 through K3 and adopted the template
star that gave the lowest mean $\chi^2$.  In both clusters, HD102494,
a G9IV star, was the ``best'' template.

Both observations and simulations by J$\o$rgensen et al. (1995b)
showed that, at low S/N, measured velocity dispersions were
systematically too large.  At the same S/N, this systematic effect was
largest for the galaxies with velocity dispersions below $100\ km\
s^{-1}$.  Therefore, we omit all galaxies with $\sigma < 100\ km\
s^{-1}$ from our samples.  Furthermore, we limit our samples to
sources with an error in $\sigma$ smaller than 15\,\%. Errors in the
dispersion and velocity are initially determined from the local
topology of the $\chi^2(\sigma,V)$ surface.

\begin {figure} [htbp]
\centering
\plotone{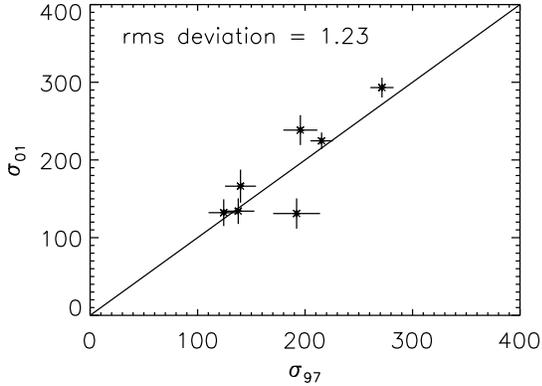}
\caption{\small
A direct comparison of velocity dispersions for 7 galaxies in
\2053.  Measured $\sigma$ values, prior to aperture correction,
for the 2001 run are plotted against $\sigma$ values from the 1997
spectra.  Formal errors derived from $\chi^2(\sigma,V)$ are drawn.
Based on this overlap sample, actual error bars are estimated to be
23\% larger.
\label {sigcomp.fig}
}
\end {figure}

Seven \2053\ galaxies with $\sigma>100\ km\ s^{-1}$ were observed both
during the 1997 and 2001 observing run.  A direct comparison between
the derived velocity dispersions (prior to the aperture correction) is
presented in Fig.\ \ref{sigcomp.fig}.  Generally, the agreement is
good, with one outlier.  All spectra have similar S/N.  The rms value
of $\frac{\sigma_{97} - \sigma_{01}}{\sqrt{d\sigma_{97}^2 +
d\sigma_{01}^2}}$ is 1.23, slightly higher than the expected value of
1.  The mean deviation between the two runs is $-2 \pm 5 \%$,
consistent with zero.  We conclude there is no evidence for systematic
effects and conservatively multiply all errors by the factor 1.23.

The final sample consists of 26 galaxies (19 early-type) in \2053\ and
22 galaxies (12 early-type) in \1054.

We applied an aperture correction to a nominal aperture of $D =
3\farcs 4$ at the distance of Coma (see JFK96).  The final velocity
dispersions have therefore been multiplied by a factor of 1.057 for
\2053 and 1.062 for \1054.  This correction allows for a fair
comparison between clusters at a range of redshifts.  The data are
tabulated in Table\ \ref{ms2053andms1054.tab}.

\section {IMAGING}
\label {imaging.sec}

For both \2053\ and \1054\ large HST WFPC2 mosaics were
obtained, each consisting of 6 pointings.  Both clusters were observed
in the F606W and F814W filters.  The layout of the \2053\ mosaic is
described in Hoekstra et al.\ (2002).  Exposure times were 3300\,s in
F606W and 3200\,s in F814W per pointing.  The \1054\ mosaic is described in van
Dokkum et al.\ (2000); exposure times were 6500\,s for each
pointing and in each filter.
Interlacing of the images improved the sampling by a factor $\sqrt{2}$
for \1054.

\subsection {Structural parameters}

In this section, we describe the measurement of effective radii $r_e$
and surface brightnesses at those radii $I_e$.  For the Johnson B
passband, $I_e$ in $L_{\sun}\ pc^{-2}$ is related to $\mu_e$ in $mag\
arcsec^{-2}$ as
\begin{equation}
\log I_e = -0.4(\mu_e - 27.0).
\end{equation}
We created postage
stamps sized $12\farcs 8 \times 12\farcs 8$ for the 26 \2053\ and 22
\1054\ galaxies and fit 2D $r^{1/n} (n=1,2,3,4)$ law profiles,
convolved with Point Spread Functions (PSF), to the galaxy images.
As PSFs depend on the positions of objects on the CCDs, we used Tiny
Tim v6.0 to create an appropriate PSF for each galaxy.  Other
parameters determining the shape of the PSF are template spectrum (M
type star), PSF size ($3\arcsec$), sampling and filter (F814W).  The
code allows simultaneous fitting of the object of interest and any
neighbouring objects. The fits were restricted to radii of $3\arcsec$
to $5\arcsec$ around the objects, depending on their size.  Image
defects were masked in the fit, as well as neighbouring galaxies not
well fitted by $r^{1/4}$ laws.  All other pixels got uniform weight.

\begin {figure*}[htbp]
\centering
\plotone{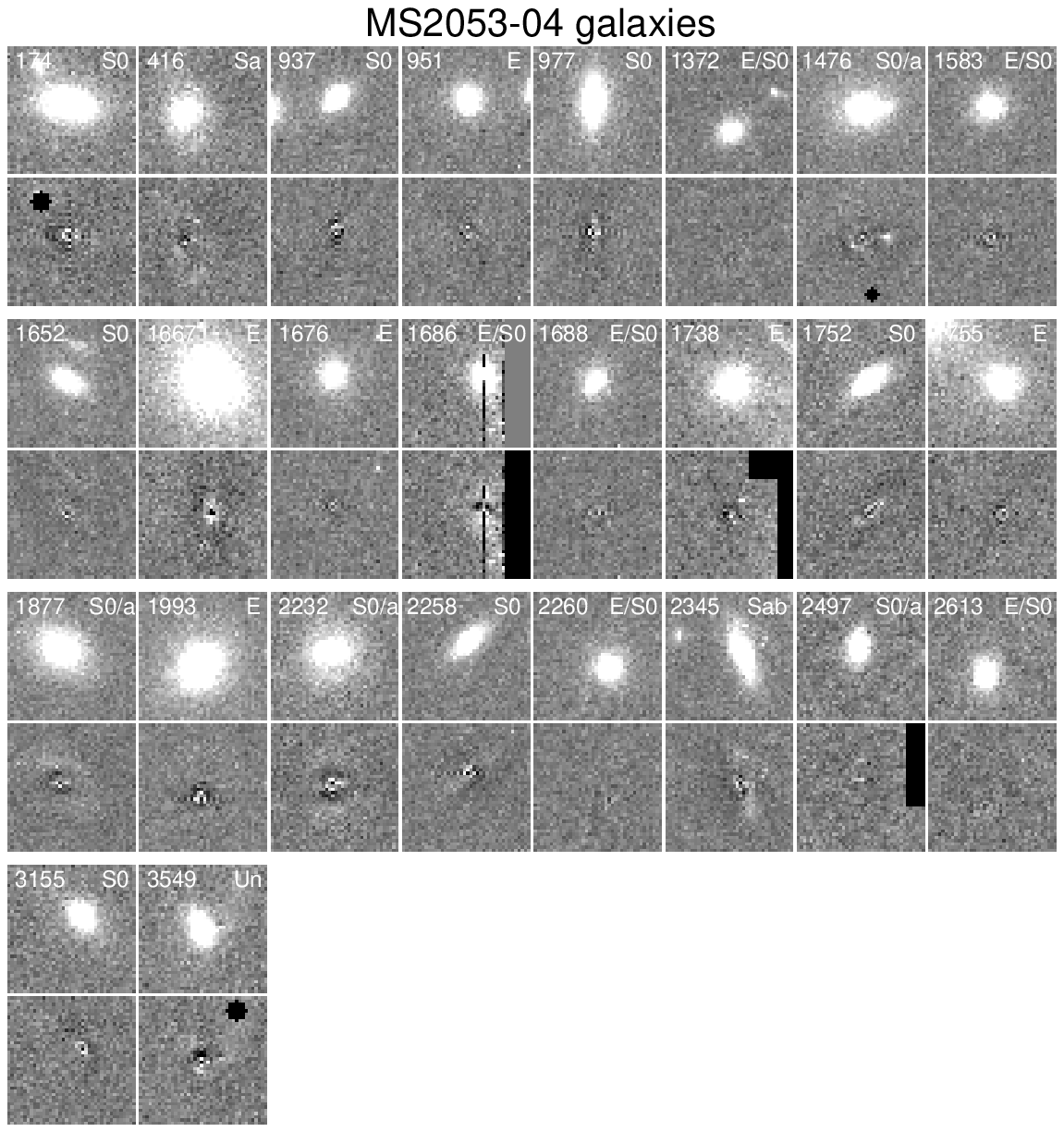}
\caption{\small $4\arcsec \times 4\arcsec$ images (upper) and
residuals (lower) after $r^{1/4}$ profile fitting for the galaxies in
\2053.  Masked regions are indicated in black.  The postage stamps
prove that our results for the early-type galaxies are not suffering
from misclassifications or bad profile fitting.
\label {mozaiek.fig}
}
\end{figure*}

\begin {figure*}[htbp]
\centering
\plotone{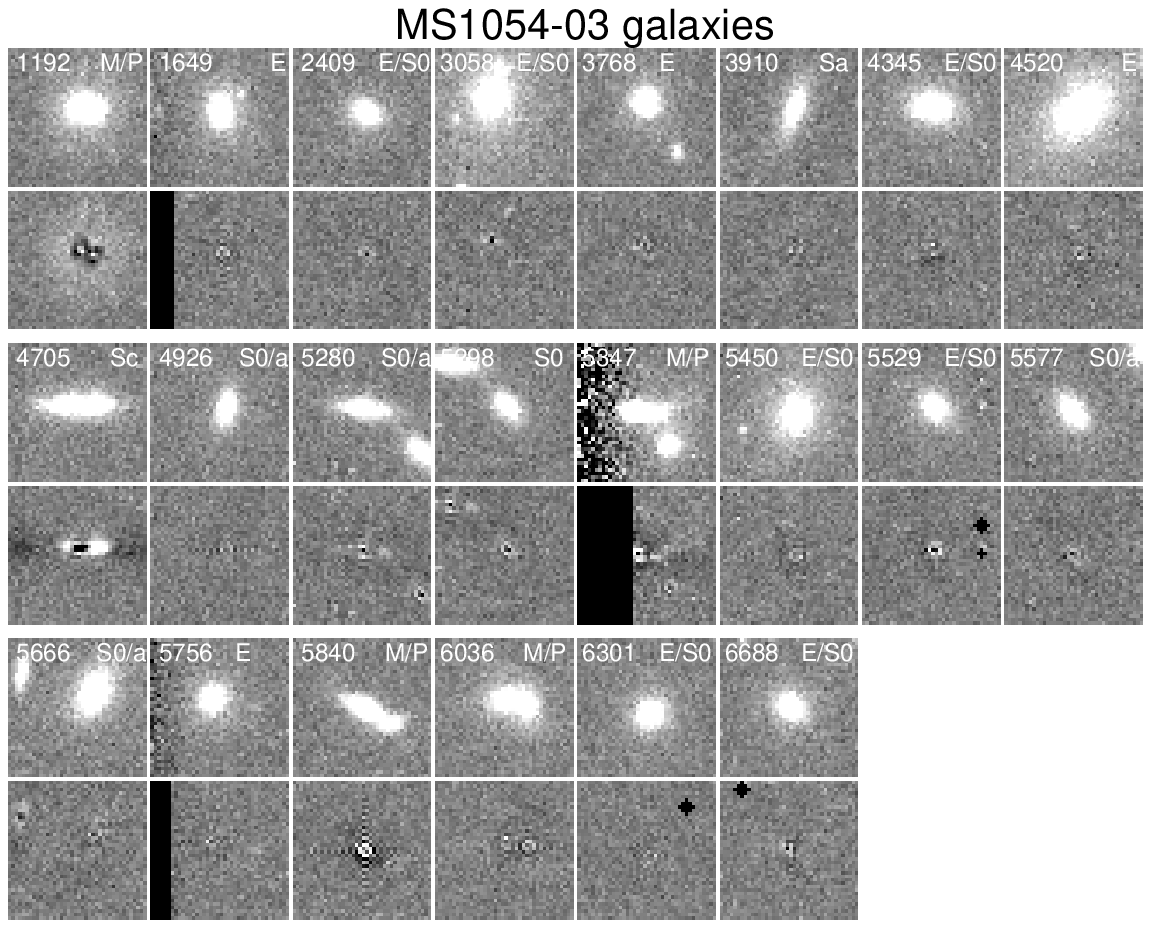}
\caption{\small $4\arcsec \times 4\arcsec$ images (upper) and
residuals (lower) after $r^{1/4}$ profile fitting for the galaxies in
\1054.  Masked regions are indicated in black.  The postage stamps
prove that our results for the early-type galaxies are not suffering
from misclassifications or bad profile fitting.
\label {mozaiek2.fig}
}
\end{figure*}

We performed the $r^{1/n}$ fits for Sersic numbers n = 1
(exponential), 2, 3 and 4 (de Vaucouleurs law).  In this paper we
always use $r_e$ and $I_e$ based on $r^{1/4}$ fits to all galaxies on
the postage stamps, even if other Sersic numbers result in a better
$\chi ^2$ of the fit.  Fitting a $r^{1/4}$ profile resulted in a $\chi
^2<1.5$ for 69\% of all galaxies; 86\% have $\chi ^2<2$.  The galaxies
in our final samples together with the residuals after profile fitting
are presented in Fig.\ \ref{mozaiek.fig} (\2053) and Fig.\
\ref{mozaiek2.fig}(\1054).

\subsection {Error in the structural parameters}

For \1054\ each of the 6 pointings was observed twice, with a shift of
0.5 pixels, providing a direct way to measure the error in the
structural parameters.  From fits to the two independent observations
we infer that errors are small ($<4\%$ in $r_e$) for galaxies larger
than $0\farcs 83$.  The error in $r_e$ rises to at most $18.5\%$ for
smaller sources.  Though this might seem problematic, we note that
this large error does not enter the FP analysis as the combination
$r_e I_e^{-\beta}$ enters the FP.  Using $\beta = -0.83$ from JFK96,
the error in the FP parameter $r_e I_e ^{0.83}$ is limited to $\sim
2.1\%$ rms, ignoring one outlier with $30\%$ offset.  The fit of a de
Vaucouleurs profile to the outlier ID5347 with merger/peculiar
morphology is clearly unstable.  The uncertainty in the FP parameter
is comparable to the $\sim 2.5\%$ rms error estimate from Kelson et
al. (2000a).  The small error in the FP parameter is an artifact of
the slope of the de Vaucouleurs growth curve.  Hereafter, we use the
average of the independent $r_e$ measurements (which reduces the error
by $1/\sqrt{2}$) and the $I_e$ corresponding to this average value,
using the empirical result that $r_e I_e^{0.66}$ is the most stable
combination of the structural parameters.  In general we can conclude
from the error analysis that random errors are small.  Therefore the
scatter found around the FP will not be due to errors in the
photometry.

\subsection {Visual and quantitative classifications}
\label{classification.sec}

Galaxies were visually classified by P.G. van Dokkum, M. Franx,\&
D. Fabricant using the procedure as described in Fabricant, Franx,\&
van Dokkum (2000).  In this paper we consider cluster members with
early-type morphology (E, E/S0, S0).  Late-type morphologies are also
plotted, but are not included in fitting procedures unless mentionned
otherwise.  A quantitative alternative to the classification by eye is
based on the Sersic number that results in the smallest $\chi ^2$ of
the profile fit.

\begin {figure} [htbp]
\centering
\plotone{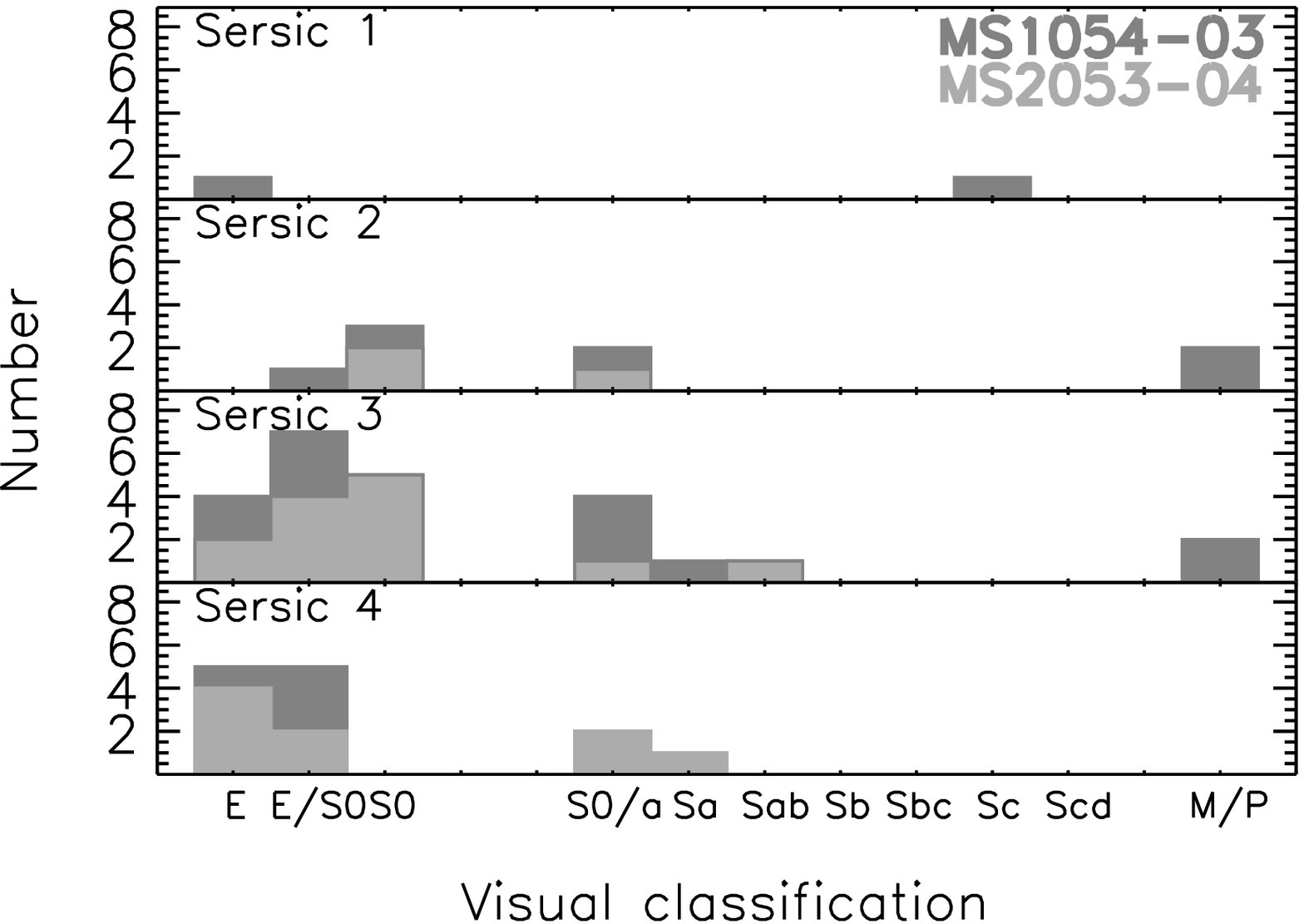}
\caption{\small
Visual classification histogram for different best fitting Sersic
numbers.  Early-type morphologies generally correspond to Sersic
numbers 3 and 4.  A different greyscale is used for \2053\ and
\1054.
\label {visSersic.fig}
}
\end {figure}

Fig.\ \ref{visSersic.fig} shows histograms of the visual
classifications for the galaxies with best fitting Sersic number 1, 2,
3 and 4.  As could be expected, early-type morphologies generally
correspond to Sersic numbers 3 and 4.  We note that
the difference in $\chi ^2$ between n = 3 and 4 is often too small to
consider them as a seperate class of objects.  We conclude that the
visual and quantitative classifications divide the sample in roughly
the same bulge and disk dominated classes.

\subsection {Transformation to restframe magnitude}

In order to compare the FP of \2053\ and \1054\ with the FP of
Coma at $z=0.023$ we have to transform the effective radius to units of kpc.
Furthermore, for a
meaningful comparison it is necessary to study all data in a common
photometric band in the restframe of the galaxies.
The observed F606W and F814W filters straddle the redshifted $B_z$
band for $0.5\leq z\leq 0.8$. Therefore, we
transform the observed
surface brightnesses to rest-frame $B$.  The
procedure is described in van Dokkum \& Franx (1996). It 
involves an interpolation between
F606W and F814W, and is different from applying a ``$K$-correction''.
For $z=0.583$ we find
\begin {equation}
\mu_{B_z}=\mu_{F814W}+0.42(F606W-F814W)+0.84,
\label {Bz2053.eq}
\end {equation}
and for $z=0.83$
\begin {equation}
\mu_{B_z}=\mu_{F814W}+0.01(F606W-F814W)+1.13.
\label {Bz1054.eq}
\end {equation}

The $F606W-F814W$ colors in (\ref {Bz2053.eq}) and (\ref {Bz1054.eq})
are obtained using SExtractor (Bertin \& Arnouts 1996) with fixed
apertures of $0\farcs 7$ diameter.  Extinction corrections for both
fixed aperture colors and surface brightnesses were derived from
Schlegel et al. (1998).  Another correction compensates for
cosmological dimming $\propto (1+z)^4$.  The final samples are
summarized in Table\ \ref{ms2053andms1054.tab}.  Included are the
coordinates with respect to the BCG, aperture corrected $\sigma$,
$r_e$, $\mu_{B_z}$ corrected for galactic extinction and cosmological
dimming, total $F814W^T$ magnitudes and morphological classifications.

\section {THE FUNDAMENTAL PLANE}
\label {jorg.sec}

\begin{figure*}[htb]
\centering
\plotone{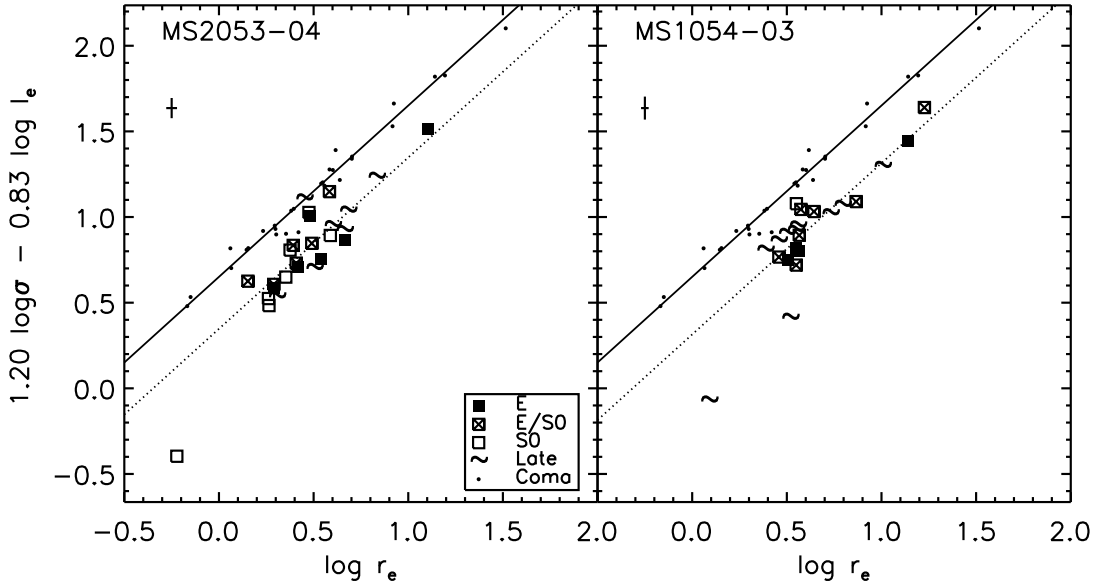}
\caption{ The fundamental plane of clusters \2053\ (z=0.583) and
\1054\ (z=0.83).  The Coma FP is drawn for reference.  Typical error
bars are plotted in the upper left corner.  Cluster galaxies in the
higher redshift clusters follow the FP scaling relation, but with an
offset with respect to Coma.  Galaxies with early-type morphologies
show a larger scatter than in the local universe.
\label{edge_on.fig}
}
\end{figure*}

In this section we discuss the FP relation in \2053\ and
\1054.  For determination of the zeropoint and scatter
we adopt the slope $(\alpha,\beta) =
(1.20,-0.83)$ that JFK96 found for the local B-band FP.  In Fig.\
\ref{edge_on.fig} we show the edge-on view of the FP for both
clusters.  Different symbols indicate different morphological types.
For comparison the Coma FP is drawn as well.  The galaxies in
\2053\ and \1054\ follow a similar FP, but with an offset with
respect to Coma.

\subsection {Zeropoint and scatter}
\label{ZPscat.sec}

\subsubsection {\2053}
\label{ZPscat2053.sec}

Using the biweight mean (Beers et al. 1990) we fit a zeropoint of the
FP to the early-type galaxies in \2053.  Under the assumption of
homology, the zeropoint shift of the FP traces the mean evolution of
the galaxy $M/L$ ratio.  The observed zeropoint offset is $\Delta \log
(M/L_B)=-0.365 \pm 0.037$, larger than $\Delta \log (M/L_B)=-0.280 \pm
0.036$ found by K97 based on older data for a sample of 5 bright
galaxies.  We will return to this issue in Sect.\
\ref{correlation.sec}.

We find a biweight scatter for the early-type population in \2053\
as large as $0.134 \pm 0.034$ in $\log r_e$, with the error derived
from bootstrapping.  This is significantly larger than the observed
scatter of 0.071 around the B-band FP of local clusters (JFK96).  After
subtraction of the measurement uncertainties in quadrature, we find
the intrinsic scatter for the early-type galaxies to be $0.124 \pm
0.037$.  We conclude that measurement uncertainties cannot account for
all of the enhanced scatter.  Not only is the scatter larger than in
the local universe, it also exceeds the previous result of $0.058 \pm
0.018$ obtained by K97.  Our new measurements for 4 early-type
galaxies that were also in the K97 sample give a scatter of $0.116 \pm
0.035$ (as opposed to $0.050 \pm 0.018$ for the original K97 data on
these 4 galaxies).  A larger sample and new data on previously studied
objects leads to the conclusion that the early-type galaxies in \2053\ show
a considerably larger spread around the FP than early-type galaxies in the
local universe.

We next analyzed the scatter of the bulge-dominated systems selected
by Sersic index (3 and 4), which is a more objective method of
classifying.  The scatter drops by about 20\%  with respect to the
classification by eye, to $0.111 \pm 0.024$.  Using both visual and
quantitative classifications, we find that bulge-dominated systems in
\2053\ are less tightly confined to a plane than in the local
universe by a factor 1.5 to 2.

Zeropoint shifts and scatters from biweight statistics are summarized
in Table\ \ref{ZPscatter.tab}.  Median zeropoint shifts are given
for comparison.

\subsubsection {\1054}
\label{ZPscat1054.sec}

For the early-type galaxies in \1054\ the zeropoint offset agrees
well with the previous result from vD98 ($\Delta \log
(M/L_B)=-0.405 \pm 0.037$ for the new sample and $\Delta \log (M/L_B)=-0.393
\pm 0.040$ from vD98).  The observed scatter is $0.106 \pm 0.024$,
and the intrinsic scatter is $0.086 \pm 0.028$,
consistent with the scatter in local clusters (JFK96).  Using new
measurements of 5 \1054\ early-type galaxies studied by vD98, we
obtain a biweight scatter of $0.062 \pm 0.018$, consistent with $0.047
\pm 0.024$ for the original vD98 data on these objects.  As for
\2053, the scatter decreases by roughly 20\% if we select
bulge-dominated systems by Sersic index (3 and 4) instead of by eye.

\begin{deluxetable*}{lrrrr} 
\tablecolumns{5}
\tablewidth{0pc}
\tablecaption{\2053\ and \1054\ zeropoint and scatter around FP.  Earlier results from K97 and vD98 are also in this Table. \label {ZPscatter.tab}} 
\tablehead{
\colhead{Sample} & \colhead{\# objects} & \colhead{$\Delta \log (M/L_B)$} & \colhead{$\Delta \log (M/L_B)$} & \colhead{Scatter
in $\log r_e$} \\
\colhead{} & \colhead{} & \colhead{(biweight mean)} & \colhead{(median)} & \colhead{(biweight)}
}
\startdata 
\2053\ early-type & 19 & $-0.365 \pm 0.037$ & $-0.404 \pm 0.037$ & $0.134 \pm 0.034$ \\
\2053\ Sersic34 & 23 & $-0.382 \pm 0.028$ & $-0.404 \pm 0.028$ & $0.111 \pm 0.024$ \\
\2053\ K97 & 5 & $-0.280 \pm 0.036$ & $-0.257 \pm 0.036$ & $0.058 \pm 0.018$ \\
\tableline
\1054\ early-type & 12 & $-0.405 \pm 0.037$ & $-0.418 \pm 0.037$ & $0.106 \pm 0.023$ \\
\1054\ Sersic34 & 15 & $-0.368 \pm 0.027$ & $-0.392 \pm 0.027$ & $0.086 \pm 0.026$ \\
\1054\ vD98 & 6 & $-0.393 \pm 0.040$ & $-0.405 \pm 0.040$ & $0.049 \pm 0.018$
\enddata
\end{deluxetable*}

\section{CORRELATIONS WITH OTHER PARAMETERS}
\label {correlation.sec}

\begin{figure*}[htbp]
\centering
\epsscale{0.9}
\plotone{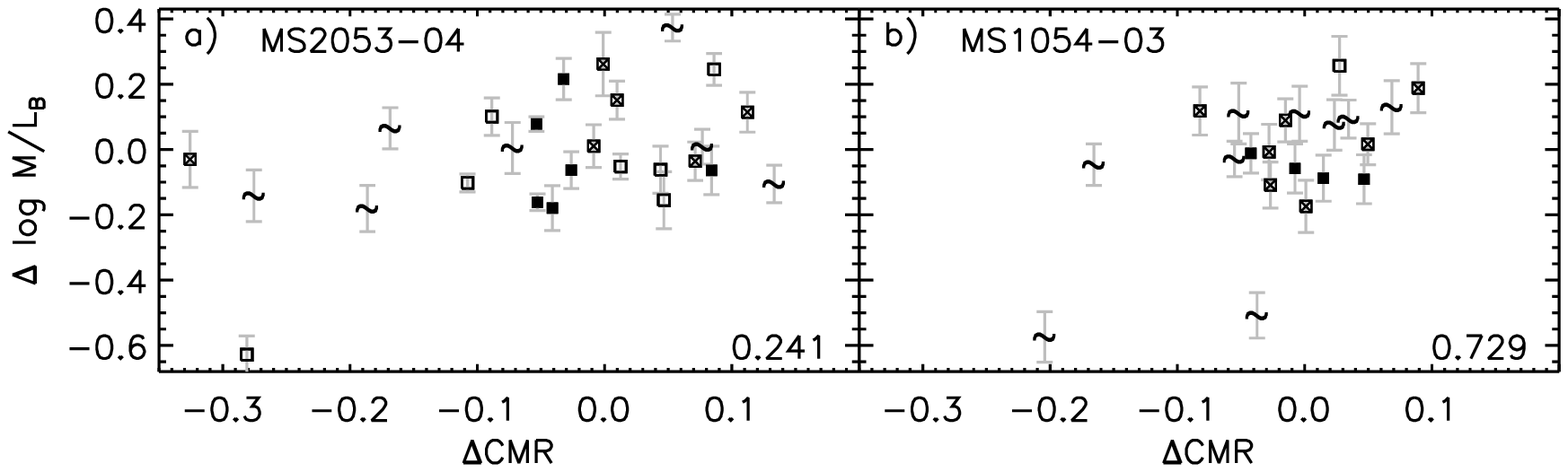}
\plotone{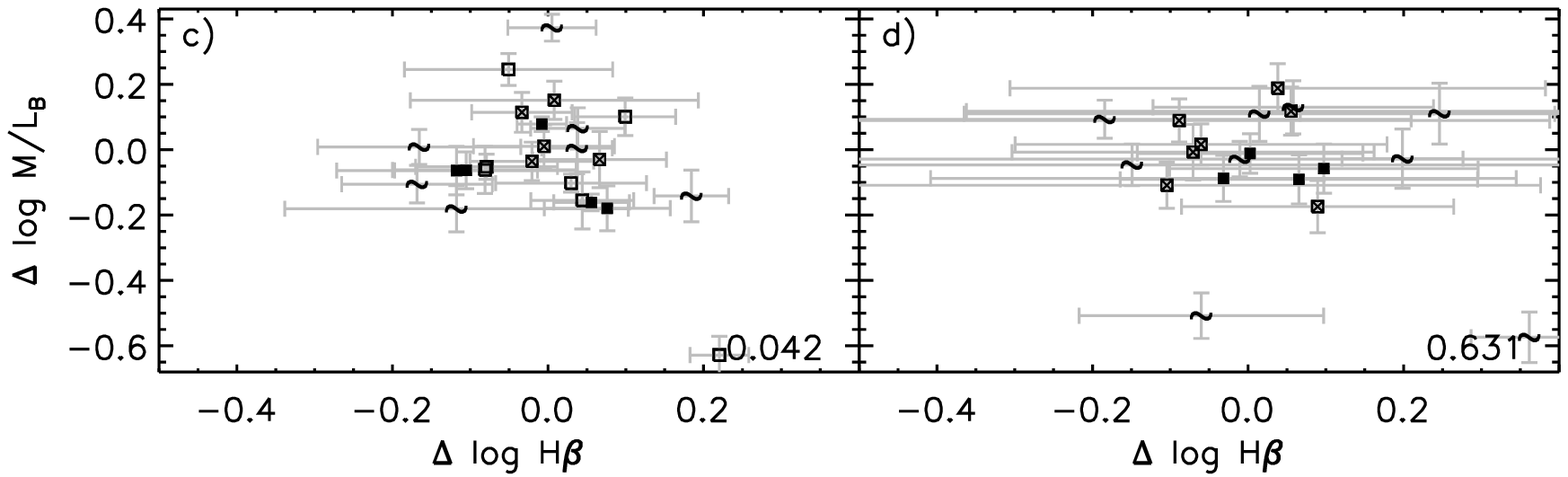}
\plotone{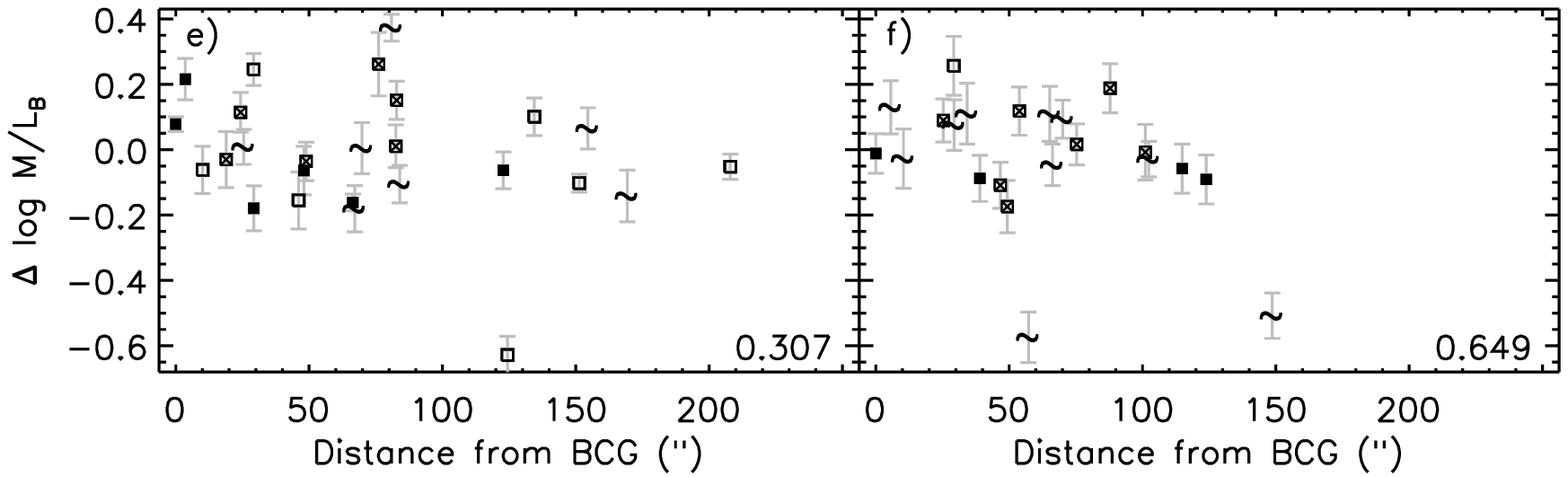}
\plotone{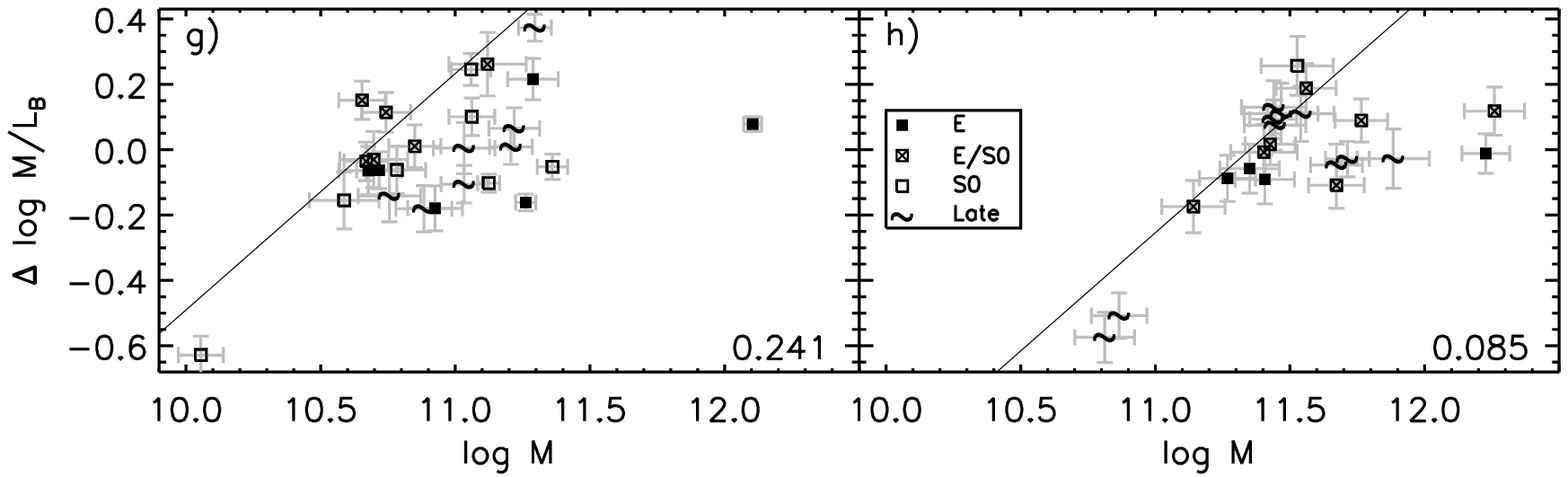}
\caption{ Residual from the fundamental plane $\Delta \log (M/L_B)$
plotted against residual from the color-magnitude relation $\Delta
CMR$, residual from the $H\beta - \sigma$ relation, distance from BCG
and galaxy mass.  Different symbols refer to different morphological
types.  Error bars for colors and angular distances are assumed to be
smaller than the symbol sizes.  For the early-type galaxies, the
p-values for statistical significance from the Spearman rank order
correlation test are printed in the lower right corners.  In panels g)
and h) the magnitude limit $I\approx 21.15$ at which serious
incompleteness due to uncertainties in the $\sigma$ measurements sets
in, is indicated with the solid line.
\label{corrtot.fig}
}
\end{figure*}

One of the striking effects we found in Sect.\ \ref{ZPscat.sec} is the
large FP scatter for \2053.  Here we investigate the cause of this
enhanced scatter.  In the case of a stellar population effect such as
variations in age, metallicities or dust content, we expect the
residual from the FP to correlate with color and linestrength indices.
We also discuss the residual from the FP as a function of environment
and investigate the dependence on galaxy mass.  Hereafter, we again
adopt the JFK96 slope and we express residuals from the FP as
deviations in $\log (M/L_B)$.  A positive residual means the galaxy
has a higher $M/L$ than the FP prediction based on its $r_e$ and
$\sigma$.  The approach of measuring offsets along the surface
brightness axis is physically intuitive since -if we ignore mergers- a
galaxy only moves along this axis during its lifetime.  For
consistency with our FP analysis, we adopt locally determined slopes
for all other considered scaling relations as well.  In Fig.\
\ref{corrtot.fig} the residual from the FP is plotted against various
properties for both clusters.  Different symbols refer to different
morphological classes.  For each panel, the probability that a random
sample has the same Spearman rank order correlation coefficient as the
early-type galaxies in our sample, is printed in the corner.

\subsection{The color-magnitude relation}
\label {cmr.sec}

The most straightforward interpretation for the large scatter is a
stellar population effect.  For age, metallicity, and dust trends, we
expect lower $M/L$ ratios than the FP prediction to correlate with
bluer colors than the CMR prediction.  We refer to Tran et al.\ (2003)
for the color-magnitude relation of \2053\ and to van Dokkum et al.\
(2000) for the \1054\ CMR.  Similar to our FP analysis, we assume that
the slope of the relation does not evolve and adopt the slope measured
in the Coma cluster by Bower, Lucey,\& Ellis (1992b).  After
conversion of the (F606W,F814W) photometry to Johnson U and V (using
the same procedure as K2000), we fit a CMR zeropoint to the early-type
galaxies in our samples.  A positive residual $\Delta CMR$ in Fig.\
\ref{corrtot.fig}a and Fig.\ \ref{corrtot.fig}b corresponds to a
redder color of the galaxy than the CMR prediction based on its total
V magnitude.  The Spearman rank order correlation coefficient points
in the expected direction for a stellar population effect.  However,
the level of significance is insufficient to confirm that galaxies
with lower $M/L$ than the FP prediction are bluer and the higher (more
evolved) ones are redder than the CMR prediction.

\subsection{$H\beta$ linestrength}
\label {linestrength.sec}

Apart from $M/L$ ratio (residual from FP) and color
(residual from CMR), strengths of absorption lines are a valuable
tool for tracing stellar population effects.  In this section we
discuss the $H\beta$ index (expressed in $\AA$, see Trager et al. 1998).  $H\beta$ is
an age-sensitive parameter with only a minor contribution due to
metallicity.  For spectral reduction and derivation of linestrengths
we refer to Kelson et al. (2003, in preparation). 

The $H\beta - \sigma$ relation for early-type galaxies in the Coma
cluster was derived from the $H\beta_G - \sigma$ relation presented by
J$\o$rgensen (1999).  The $H\beta_G$ index is related to the Lick/IDS
$H\beta$ index as $H\beta_G=0.866 H\beta+ 0.485$ (J$\o$rgensen 1997).
Assuming a non-evolving slope of the $H\beta - \sigma$ scaling
relation, we fit a zeropoint to the relation in \2053\ and \1054.  The
residuals from this relation (positive means stronger $H\beta$) are
plotted against the FP residual in Fig.\ \ref{corrtot.fig}c and Fig.\
\ref{corrtot.fig}d.  Only for \2053\ are the error bars small enough
to draw robust conclusions.  A correlation is present, with confidence
level 95.8\%.  $H\beta$ absorption lines are stronger for younger
stellar populations, and the correlation with $\Delta \log (M/L_B)$
could confirm that the scatter in the FP is not random noise, but
determined by age variations among the galaxies.

\subsection{Location in the cluster}
\label {location.sec}

Here we consider if the enhanced scatter reported in Sect.\
\ref{ZPscat.sec} is related to environment.  Clusters of galaxies are
not isolated systems, as infall of galaxies from the field occurs.  FP
studies of field early-type galaxies indicate that their stellar
populations may be somewhat younger than those of their counterparts
in clusters (van Dokkum et al.\ 2001; Treu et al. 2002; van de Ven,
van Dokkum,\& Franx 2003; Rusin et al. 2003; van Dokkum \& Ellis 2003;
Gebhardt et al.\ 2003).  In the cluster CL1358+62 van Dokkum et al.\
(1998) reported evidence for disky galaxies to be systematically bluer
and to show a larger scatter in the CMR at larger radii (with the
sample ranging to $\sim 1 Mpc$).  As earlier studies of both clusters
(K97; vD98) were based on one pointing, the larger scatter we find
could be explained if residuals from the FP increase with distance
from the BCG.  Therefore, we might expect to see a gradient in age and
larger scatter around the FP going to a larger range in cluster radii.
For both clusters our samples extend to roughly $1 Mpc$ from the BCG.
Fig.\ \ref{corrtot.fig}e and Fig.\ \ref{corrtot.fig}f do not show a
significant correlation.

\subsection{Galaxy mass and selection effects}
\label {mass.sec}

Finally we try to explain the range of FP residuals as a function of
galaxy mass.  The mass $M$ in solar units of a galaxy is calculated as
follows (see JFK96):
\begin{equation}
\log M = 2\log \sigma + \log r_e + 6.07
\end{equation}
Fig.\ \ref{corrtot.fig}g and Fig.\ \ref{corrtot.fig}h show residual
from the FP against $\log M$.  Only if we were to include the 6
galaxies with $\sigma < 100\ km\ s^{-1}$, the trend of lower mass
galaxies to have lower $M/L$ ratios than predicted by the FP with
JFK96 coefficients is significant at the 95\% level.\\ 

For a good understanding of Fig.\ \ref{corrtot.fig}g and Fig.\
\ref{corrtot.fig}h, we need to take into account that our FP samples
are magnitude-limited, and not mass-limited.  The FP can be rewritten
as
\begin {equation}
M/L \sim M^{0.28}r_e^{-0.07}.
\end {equation}
In the following, we ignore the dependence on $r_e$.  Hence the residual is given by
\begin {equation}
\Delta M/L = \frac{M_{obs}/L_{obs}}{(M/L)_{FP}} = \frac{M_{obs}^{0.72}}{L_{obs}}.
\end {equation}
For a fixed luminosity we expect all galaxies to fall on a line in a
plot of $\Delta \log M/L$ versus $\log M$.  In Fig.\
\ref{corrtot.fig}g and Fig.\ \ref{corrtot.fig}h this line is drawn for
I=21.15.  At this magnitude serious incompleteness due to
uncertainties in the $\sigma$ measurements sets in.  The lowest mass
galaxies in Fig.\ \ref{corrtot.fig}a and Fig.\ \ref{corrtot.fig}b lie
close to the line representing the magnitude limit.  Low mass galaxies
that lie on or above the FP would be too faint to allow for accurate
dispersion measurements and would not enter the FP samples.  Hence the
few remaining low mass galaxies in our samples are brighter than their
FP prediction based on $r_e$ and $\sigma$.\\ 

Selection effects are clearly less relevant at the high mass end, and
hence we determine the offset and scatter of the \2053\ FP seperately
for the subsample of early-type galaxies with $M > 10^{11}M_{\sun}$
(following van Dokkum \& Stanford 2003).  For this subsample of 8
early-type galaxies we derive a zeropoint shift with respect to Coma
of $\Delta \log (M/L_B) = -0.288 \pm 0.056$ and a scatter of $0.132
\pm 0.039$ in $\log r_e$.  As expected, this is slightly different
from the zeropoint shift of $\Delta \log (M/L_B) = -0.365 \pm 0.037$
for the full early-type sample.  The zeropoint offset is now in good
agreement with K97, but the scatter is still larger.

If we apply a mass cut for the \1054\ early-type galaxies of
$M>10^{11.5}M_{\sun}$, we find a shift in zeropoint of $\Delta \log
(M/L_B) = -0.311 \pm 0.051$ compared to $\Delta \log (M/L_B) = -0.405
\pm 0.037$ for the full early-type sample.  The scatter of the high
mass subsample ($0.104 \pm 0.030$ in $\log r_e$) is similar to that of
the full early-type sample ($0.106 \pm 0.023$ in $\log r_e$).

\subsection{Summary of correlations}
\label{corrsummary.sec}

We did not find evidence that FP residuals are related to environment.
Instead, stellar population effects are playing a role in shaping the
FP and selection effects in our magnitude-limited samples need to be
taken into account.

First, early-type galaxies with stronger $H\beta$ absorption also tend
to have lower $M/L$ ratios than predicted by their $r_e$ and $\sigma$.
This correlation supports the interpretation of FP scatter as a
measure of age variation among early-type galaxies.  The larger FP
scatter in \2053\ therefore reflects a larger spread in relative ages
than in the local universe.  Comparison of FP residuals with residuals
from the CMR cannot confirm or rule out the presence of such a stellar
population effect.

Second, the fact that we do not see galaxies with low masses that lie
on or above the FP, may be entirely explained by selection effects.
Only low mass galaxies that are brightened by some recent star
formation enter the FP sample.  Their older counterparts are fainter
than the magnitude limit for the dispersion measurements.

\section {EVOLUTION OF $M/L$ RATIO}
\label{evol.sec}

\begin{figure*}[htb]
\centering
\plotone{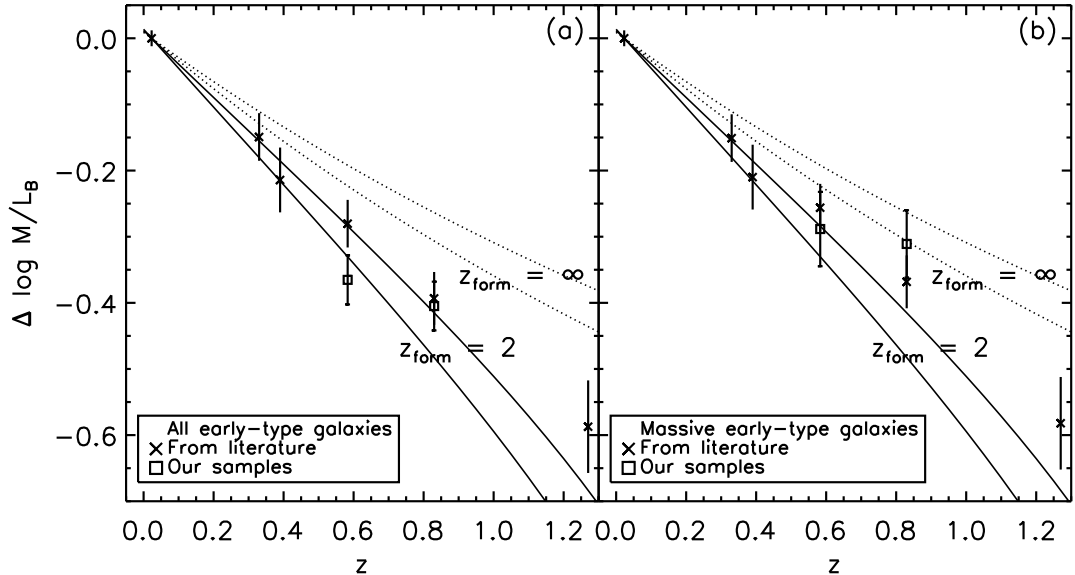}
\caption{ Evolution of the $M/L$ ratio with redshift.  The cross
symbols are results from the literature, namely Coma at $z=0.023$
(JFK96), CL1358+62 at $z=0.33$ (K2000), CL0024+16 at $z=0.39$ (van
Dokkum \& Franx 1996), \2053\ at $z=0.583$ (K97), \1054\ at $z=0.83$
(vD98) and J0848+4453 at $z=1.27$ (van Dokkum \& Stanford 2003).
Boxes denote the shift in $M/L$ ratio based on our larger samples for
\2053\ and \1054.  Single burst models for $z_{form}=2$ and $\infty$,
assuming a Salpeter IMF ($x=2.35$) and a range of metallicities, are
drawn with solid and dotted curves resp.  Panel (a) is based on all
early-type galaxies, panel (b) shows results for the subsample of
massive early-type galaxies ($M > 10^{11}M_{\sun}$ for clusters up to
$z=0.583$ and $M > 10^{11.5}M_{\sun}$ for the two higher redshift
clusters).
\label{MLevol.fig}
}
\end{figure*}

Correlations with residual from the $H\beta - \sigma$ relation show
that differences in FP zeropoint can be explained by age differences
of the stellar population.  In this section we study the evolution of
the $M/L$ ratio as a function of redshift and use this to estimate the
mean formation redshift of cluster early-type galaxies.  In Fig.\
\ref{MLevol.fig}a we show the evolution of the $M/L$ ratio with
respect to Coma. Crosses refer to measured zeropoint shifts from the
literature: Coma at $z=0.023$ (JFK96), CL1358+62 at $z=0.33$ (K2000),
CL0024+16 at $z=0.39$ (van Dokkum,\& Franx 1996), \2053\ at $z=0.583$
(K97), \1054\ at $z=0.83$ (vD98) and J0848+4453 at $z=1.27$ (van
Dokkum,\& Stanford 2003).  The new results for \2053\ and \1054, based
on our larger samples, are plotted with boxes.  Single burst
evolutionary models for a formation redshift $z_{form}=2$ and
$z_{form}=\infty$ are drawn with a solid and dotted curve
respectively.  They represent a galaxy that is fixed in mass and whose
luminosity evolves as:

\begin {equation}
L(t) \sim 1/{(t-t_{form})^{\kappa}}
\label {Levolv.eq}
\end {equation}

Here $t_{form}$ represents the age of the universe at the moment the
stars were formed.  $\kappa$ depends on the slope of the IMF, passband
and metallicity.  For a Salpeter (1955) IMF and $-0.5<[Fe/H]<0.5$
models of Bruzual \& Charlot (1998), Vazdekis et al.\ (1996) and
Worthey (1998) give $0.86< \kappa _B <1.00$.  Using the new offsets
for the two higher z clusters, the single burst model for a Salpeter
IMF and solar metallicity favoured by the least square method has
$z_{form} \sim 2.26_{-0.20}^{+0.28}$.  The $1 \sigma$ confidence level
was derived from the difference in $\chi^2$ between the model and the
overall minimum, $\Delta\chi^2 = \chi^2 - \chi^2_{min}$, to which
Gaussian confidence levels were assigned (e.g. Press et al. 1992).  As
we showed in Sect.\ \ref{mass.sec}, the point of \2053\ deviates from
the earlier result (K97) since at low mass only galaxies with low M/L
enter the magnitude-limited sample.

To avoid this bias, we apply a mass cut of $M > 10^{11}M_{\sun}$ to
all clusters up to $z=0.583$.  A mass cut of $M > 10^{11.5}M_{\sun}$
was applied to \1054\ and J0848+4453 since at these higher redshifts
the selection effect sets in at a higher galaxy mass.  For J0848+4453
two galaxies are left after omitting the one low mass outlier.  As the
biweight location estimator is robust against outliers, the zeropoint
shift only changes slightly for this cluster.  We obtain the evolution
of the $M/L$ ratio as presented in Fig.\ \ref{MLevol.fig}b.  Now the
zeropoint of the \2053\ FP follows the trend seen for the other
clusters in Fig.\ \ref{MLevol.fig}.  If we constrain the analysis of
the evolution in $M/L$ ratio to massive early-type galaxies, a simple
linear fit gives $\log M/L_B \sim -0.47z$, agreeing well with earlier
determinations based on smaller samples or lower redshift (see,
e.g. vD98).  The formation redshift favoured by a least squares method
is $z_{form} \sim 2.95_{-0.46}^{+0.81}$, slightly higher than the mean
formation redshift for all early-type galaxies.  A similar formation
epoch for early-type galaxies in clusters was found by Kelson et
al. (2001) based on the evolution of Balmer absorption-line strengths
with redshift.  It is remarkable how earlier studies of clusters based
on smaller samples (see, e.g. K97; vD98) gave similar results.

\section {SUMMARY}
\label {conclusion.sec}

We used visual and quantitative classifications to select
bulge-dominated systems in our \2053\ and \1054\ samples.  For \2053\
we find a zeropoint offset with respect to Coma of $\Delta \log
(M/L_B)=-0.365 \pm 0.037$, larger than determined earlier on the basis
of a smaller sample (K97). The scatter around the \2053\ FP is $0.134
\pm 0.034$ in $\log r_e$, enhanced with respect to both K97 and the
scatter in local clusters.  The FP zeropoint of \1054\ ($\Delta \log
(M/L_B)=-0.405 \pm 0.037$) agrees well with the earlier result of
$\Delta \log (M/L_B)=-0.393 \pm 0.040$ by vD98.  The scatter of $0.106
\pm 0.024$ in \1054\ is larger than reported by vD98 for a smaller
sample of \1054\ early-type galaxies.  However, taking into account
measurement uncertainties, the scatter is consistent with that in
local clusters (JFK96).  Late-type galaxies also follow the FP scaling
relation and show a similar scatter around the FP as the early-type
galaxies.  Adding the late-type galaxies to the early-type sample
results in a scatter of $0.136\pm0.029$ for \2053 and $0.117\pm0.031$
for \1054.  The larger samples presented in this paper allow us to
study correlations with other properties of the early-type galaxies.
We do not find evidence that the formation history depends on
environment in the cluster.  No significant correlation of FP
residuals with CMR residuals was found.  The presence of a correlation
between FP residuals and residuals from the $H\beta-\sigma$ relation
indicates that stellar population effects are playing a role.
Assuming non-evolving slopes for all scaling relations, we find that
galaxies with lower $M/L$ than the FP prediction tend to show stronger
$H\beta$ than predicted based on the $H\beta-\sigma$ relation.
Finally, we show that the lack of low mass galaxies on or above the FP
may be entirely due to selection effects.  To avoid a bias induced by
the magnitude limit of our sample, we focus on the high mass end,
where selection effects are less relevant.  Applying a mass cut at $M
> 10^{11}M_{\sun}$ to all 4 considered clusters below $z \sim 0.6$ and
at $M > 10^{11.5}M_{\sun}$ to the 2 higher redshift clusters at
$z=0.83$ and $z=1.27$, increases the best fitting formation redshift
from $z_{form}=2.26_{-0.20}^{+0.28}$ to
$z_{form}=2.95_{-0.46}^{+0.81}$.  The mass cut at $M =
10^{11}M_{\sun}$ is well below the typical mass of early-type
galaxies: galaxies with $M = 10^{11}M_{\sun}$ have dispersions of
$\sim 168\ km\ s^{-1}$ which is significantly lower than the
$\sigma_*$ dispersion of early-type galaxies which is $228 \pm 14\ km\
s^{-1}$ (Kochanek 1994).\\ The implication of this work is that
selection effects need to be taken into account, especially if the
scatter is high.  The scatter in \2053\ is slightly higher than at low
redshift; the scatter in the field at high redshift seems to be even
higher (e.g. Gebhardt et al. 2003; van Dokkum \& Ellis 2003).  Hence
those studies are likely to suffer from much more significant
selection effects.

\section {REFERENCES}
Beers, T. C., Flynn, K.,\& Gebhardt, K. 1990, AJ, 100, 32\\
Bender, R., Saglia, R. P., Ziegler, B., Belloni, P., Greggio, L., Hopp, U.,\& Bruzual, G. 1998, ApJ, 493, 529\\
Bertin, E.,\& Arnouts, S. 1996, A\&AS, 117, 393\\
Bower, R. G., Lucey, J. R.,\& Ellis, R. S. 1992b, MNRAS, 254, 589\\
Bruzual, G.,\& Charlot, S. 1998, in preparation\\
Djorgovski, S.,\& Davis, M. 1987, ApJ, 313, 59\\
Dressler, A., Lynden-Bell, D., Burstein, D., Davies, R. L., Faber, S.M., Terlevich, R. J.,\& Wegner, G. 1987, ApJ, 313, 42\\
Fabricant, D., Franx, M.,\& van Dokkum, P. G. 2000, ApJ, 539, 577\\
Gebhardt, K., et al. 2003, ApJ, in press\\
Hoekstra, H., Franx, M., Kuijken, K., van Dokkum, P. G. 2002, MNRAS, 333, 911\\
J$\o$rgensen, I., Franx, M.,\& Kj$\ae$rgaard, P. 1995b, MNRAS, 276, 1341\\
J$\o$rgensen, I., Franx, M.,\& Kj$\ae$rgaard, P. 1996, MNRAS, 280, 167 (JFK96)\\
J$\o$rgensen, I. 1997, MNRAS, 288, 161\\
J$\o$rgensen, I. 1999, MNRAS, 306, 607\\
J$\o$rgensen, I., Franx, M., Hjorth, J.,\& van Dokkum, P. G. 1999, MNRAS, 308, 833\\
Kelson, D. D., van Dokkum, P. G., Franx, M., Illingworth, G. D.,\& Fabricant, D. 1997, ApJ, 478, L13 (K97)\\
Kelson, D. D., Illingworth, G. D., Franx, M.,\& van Dokkum, P. G. 2000a, ApJ, 531, 137 \\
Kelson, D. D., Illingworth, G. D., Franx, M.,\& van Dokkum, P. G. 2000b, ApJ, 531, 159 \\
Kelson, D. D., Illingworth, G. D., Franx, M.,\& van Dokkum, P. G. 2000c, ApJ, 531, 184 (K2000)\\
Kelson, D. D., Illingworth, G. D., Franx, M.,\& van Dokkum, P. G. 2001, ApJ, 552, L17\\
Kelson, D. D. 2003, PASP, 115, 688\\
Kelson, D. D., et al. 2003, in preparation\\
Kochanek, C. S. 1994, ApJ, 436, 56 \\
Oke, J. B., et al. 1995, PASP, 107, 375\\
Press, W. H., Teukolsky, S. A., Vetterling, W. T.,\& Flannery, B. P. 1992, Numerical Recipes. Cambridge Univ. Press, Cambridge\\
Rusin, D., et al. 2003, ApJ, 587, 143\\
Salpeter, E. 1955, ApJ, 121 161\\
Schlegel, D. J., Finkbeiner, D. P.,\& Davis, M. 1998, ApJ, 500, 522\\
Trager, S. C., Worthey, G., Faber, S. M., Burstein, D.,\& Gonzalez, J. J. 1998, ApJS, 116, 1 \\
Tran, K. V., Kelson, D. D., van Dokkum, P. G., Franx, M., Illingworth, G. D.,\& Magee, D. 1999, ApJ, 522, 39\\
Tran, K. V. 2002, PhD thesis, Univ. California at Santa Cruz\\
Tran, K. V., et al. 2003, submitted to ApJ\\
Treu, T., Stiavelli, M., Casertano, S., M$\o$ller, P.,\& Bertin, G. 2002, ApJ, 564, L13\\
van de Ven, G., van Dokkum, P.G.,\& Franx, M. 2003, astro-ph/0211566 v2\\
van Dokkum, P. G.,\& Franx, M. 1996, MNRAS, 281, 985\\
van Dokkum, P. G., Franx, M., Kelson, D. D., Illingworth, G. D., Fisher, D.,\& Fabricant, D. 1998, ApJ, 500, 714\\
van Dokkum, P. G., Franx, M., Kelson, D. D.,\& Illingworth, G. D. 1998, ApJ, 504, L17 (vD98)\\
van Dokkum, P. G., Franx, M., Fabricant, D., Illingworth, G. D.,\& Kelson, D. D. 2000, ApJ, 541, 95\\
van Dokkum, P. G., Franx, M., Kelson, D. D.,\& Illingworth, G. D. 2001, ApJ, 553, 39\\
van Dokkum, P. G.,\& Stanford, S. A. 2003, ApJ, 585, 78\\ 
van Dokkum, P. G.,\& Ellis, R.S. 2003, ApJ, 592, 53\\
Vazdekis, A., et al. 1996, ApJS, 106, 307\\
Worthey, G. 1998, PASP, 110, 888W
\begin{turnpage}[hbtp]
\begin{deluxetable*}{lrrrrrrrrlrrrrrrr}
\tabletypesize{\scriptsize}
\tablewidth{0pt}
\tablecaption{FP sample \label {ms2053andms1054.tab}} 
\tablehead{
\multicolumn{8}{c}{\2053} & & \multicolumn{8}{c}{\1054} \\
\cline{1-8} \cline{10-17} \\
\colhead{ID} & \colhead{$\Delta R.A.$\tablenotemark{a}} & \colhead{$\Delta Dec$\tablenotemark{a}} & \colhead{$\log r_e$}   & \colhead{$\mu_{B_z}$\tablenotemark{b}}    & \colhead{$\sigma$} & \colhead{$F814W^T$} & \colhead{Type} & & \colhead{ID} & \colhead{$\Delta R.A.$\tablenotemark{c}} & \colhead{$\Delta Dec$\tablenotemark{c}} & \colhead{$\log r_e$}   & \colhead{$\mu_{B_z}$\tablenotemark{b}}    & \colhead{$\sigma$} & \colhead{$F814W^T$} & \colhead{Type} \\
\colhead{} & \colhead{[$\arcsec$]} & \colhead{[$\arcsec$]} & \colhead{[$kpc$]} & \colhead{[$mag\ \arcsec^{-2}$]} & \colhead{[$km/s$]} & \colhead{[mag]} & \colhead{} & & \colhead{} & \colhead{[$\arcsec$]} & \colhead{[$\arcsec$]} & \colhead{[$kpc$]} & \colhead{[$mag\ \arcsec^{-2}$]} & \colhead{[$km/s$]} & \colhead{[mag]} & \colhead{}
}
\startdata 
      174  &  9.34 & -206.60  &  0.588   &   21.19   &   $225 \pm 13$  & 19.68 & S0 &  & 1192 &  136.10 & -58.31  &  0.518 &     20.51  &    $138 \pm 15$ & 20.24 & M/P \\
      416  &  29.26 & -150.90  &  0.833   &   22.92   &   $144 \pm 14$  & 20.48 & Sa &  & 1649 &  122.30 & -16.89  &  0.565 &     20.80  &    $243 \pm 28$ & 20.69 & E \\
      937  &  -30.58 & -119.80  &  -0.223   &   18.20   &   $127 \pm 11$  & 20.70 & S0 &  & 2409 &   84.18 & -24.01  &  0.574 &     21.26  &    $287 \pm 33$ & 21.36 & E/S0 \\
      951  &  -31.74 & -118.00  &  0.288   &   20.88   &   $151 \pm 13$  & 20.81 & E &  & 3058 &  52.73 & -9.72  &  1.227 &     22.97  &    $303 \pm 33$ & 19.83 & E/S0 \\
      977  &  -72.15 & -132.10  &  0.261   &   19.92   &   $249 \pm 11$  & 20.03 & S0 &  & 3768 &  38.86 & -0.38  &  0.507 &     20.78  &    $222 \pm 24$ & 21.04 & E \\
      1372 &  -39.50 & -72.32  &  0.153  &    20.88  &   $ 164 \pm 14$ & 21.55 & E/S0 &  & 3910 &  32.00 & -11.84  &  0.457 &     20.67  &    $295 \pm 42$ & 21.23 & Sa \\
      1476 &  -38.26 & -54.80  &  0.505  &    21.33  &   $ 143 \pm 16$ & 20.04 & S0/a & & 4345 &  21.51 & -13.22  &  0.643 &     20.98  &    $336 \pm 34$ & 20.55 & E/S0 \\ 
      1583 &  -24.76 & -41.87  &  0.289  &    21.04  &   $ 143 \pm 10$ & 21.04 & E/S0 & & 4520 &  0 & 0  &  1.141 &      22.29 &    $ 322 \pm 30$ & 19.48 & E \\
      1652 &  -9.65 & -27.37  &  0.475  &    21.94  &   $ 181 \pm 13$ & 21.12 & S0 & & 4705 &  6.10 & 8.34  &  1.006 &      22.22 &    $ 253 \pm 36$ & 20.61 & Sc \\
      1667 &  0 & 0  &  1.103  &    22.65  &   $ 292 \pm 10$ & 18.56 & E & & 4926 &  -4.07 & -3.68  &  0.387  &    20.43   &   $310 \pm 38$ & 21.32 & S0/a \\
      1676 &  47.76 & 0.89  &  0.415  &    21.56  &   $ 125 \pm 14$ & 20.76 & E & & 5280 &  -20.46 & 20.82  &  0.548  &    21.06   &   $259 \pm 31$ & 21.17 & S0/a \\
      1686 &  14.28 & -12.29  &  0.408  &    21.58  &   $ 129 \pm 15$ & 20.95 & E/S0 & & 5298 &  -21.74 & 19.24  &  0.550  &    21.38   &   $284 \pm 39$ & 21.52 & S0 \\
      1688 &  -9.00 & -22.49  &  0.392  &    21.78  &   $ 138 \pm 13$ & 21.31 & E/S0 & & 5347 &  -34.24 & 56.37  &  0.795   &   21.54    &  $254 \pm 24$ & 20.58 & M/P \\
      1738 &  -23.00 & -17.84  &  0.667  &    22.04  &   $ 124 \pm 13$  & 20.01 & E & & 5450 &  -46.30 & -3.51  &  0.866  &    21.72   &   $234 \pm 26$ & 19.98 & E/S0 \\
      1752 &  -1.54 & -9.86  &  0.353  &    21.08  &   $ 151 \pm 17$ & 20.88 & S0 & & 5529 &  -37.79 & 31.29  &  0.549  &    20.99   &   $182 \pm 23$ & 21.01 & E/S0 \\
      1755 &  -1.14 & -3.41  &  0.482  &    21.48  &   $ 234 \pm 23$ & 20.23 & E & & 5577 &  -43.54 & 48.16  &  0.501   &   20.75    &  $305 \pm 40$ & 21.05 & S0/a \\
      1877 &  19.96 & 15.62  &  0.683  &    22.07  &   $ 169 \pm 14$ & 20.17 & S0/a & & 5666 &  -58.07 & -83.77  &  0.731   &   21.20    &  $286 \pm 23$ & 20.50 & S0/a \\
      1993 &  49.67 & 43.54  &  0.540  &    20.86  &   $ 212 \pm 8$ & 19.40 & E & & 5756 &  -59.26 & 97.86  &  0.551   &   20.91    &  $232 \pm 27$ & 20.93 & E \\
      2232 &  -82.50 & 13.16  &  0.664  &    22.01  &   $ 141 \pm 12$ & 20.16 & S0/a & & 5840 &  -52.17 & 23.03  &  0.090    &  18.38     & $212 \pm 26$ & 20.66 & M/P \\  
      2258 &  -2.50 & 45.70  &  0.264  &    20.77  &   $ 134 \pm 18$ & 20.93 & S0 & & 6036 &  -63.37 & -29.09  &  0.559  &    21.17   &  $ 254 \pm 22$ & 21.18 & M/P \\ 
      2260 &  46.61 & 67.51  &  0.490  &    21.80  &   $ 140 \pm 14$ & 20.81 & E/S0 & & 6301 &  -70.57 & 25.38  &  0.565  &    21.03   &  $ 249 \pm 24$ & 20.99 & E/S0 \\ 
      2345 &  16.82 & 67.38  &  0.601  &    21.99  &   $ 152 \pm 19$ & 20.58 & Sab & & 6688 &  -91.54 & -41.79  &  0.458  &    20.50   &   $274 \pm 37$ & 20.82 & E/S0 \\
      2497 &  15.46 & 78.88  &  0.451  &    21.71  &   $ 244 \pm 14$ & 21.13 & S0/a & & & & & & & & & \\
      2613 &  -23.80 & 71.80  &  0.584  &    22.39  &   $ 171 \pm 25$ & 21.07 & E/S0 & & & & & & & & & \\
      3155 &  -90.37 & 98.46  &  0.376  &    21.09  &   $ 203 \pm 18$ & 20.64 & S0 & & & & & & & & & \\
      3549 &  -29.45 & 165.70  &  0.307  &    20.70  &   $ 155 \pm 19$ & 20.45 & Un & & & & & & & & & \\
\enddata
\tablenotetext{a}{Coordinates with respect to the BCG of \2053: ID1667 at (20:56:21.4; -04:37:50.8) (J2000).}
\tablenotetext{b}{Surface brightnesses $\mu_{B_z}$ are corrected for galactic extinction and cosmological dimming.}
\tablenotetext{c}{Coordinates with respect to the BCG of \1054: ID4520 at (10:56:59.9; -03:37:37.3) (J2000).}
\end{deluxetable*}
\end{turnpage}
\end {document}